\documentclass{article}

\usepackage{arxiv}

\usepackage[utf8]{inputenc} % allow utf-8 input
\usepackage[T1]{fontenc}    % use 8-bit T1 fonts
\usepackage{hyperref}       % hyperlinks
\usepackage{url}            % simple URL typesetting
\usepackage{booktabs}       % professional-quality tables
\usepackage{amsfonts}       % blackboard math symbols
\usepackage{nicefrac}       % compact symbols for 1/2, etc.
\usepackage{microtype}      % microtypography
\usepackage{lipsum}		% Can be removed after putting your text content
\usepackage{graphicx}
\usepackage{natbib}
\usepackage{doi}

\title{From Interests to Insights: An LLM Approach to Course Recommendations Using Natural Language Queries}

\author{ \href{https://orcid.org/0009-0008-4851-885X}{\includegraphics[scale=0.06]{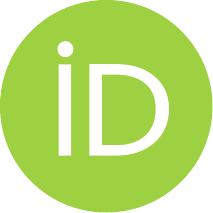}\hspace{1mm}Hugh Van Deventer}\thanks{\href{https://hughvd.github.io/}{https://hughvd.github.io}} \\
	LSA \\
	University of Michigan\\
	New York, NY \\
	\texttt{hughv@umich.edu} \\
	\And
	\href{https://orcid.org/0000-0002-1145-9592}{\includegraphics[scale=0.06]{orcid.pdf}\hspace{1mm}Mark Mills} \\
	Center for Academic Innovation\\
	University of Michigan\\
	Ann Arbor, MI \\
	\texttt{marmills@umich.edu} \\
        \And
	\href{https://orcid.org/0000-0002-4876-956X}{\includegraphics[scale=0.06]{orcid.pdf}\hspace{1mm}August E.~Evrard} \\
	Department of Physics\\
	University of Michigan\\
	Ann Arbor, MI \\
	\texttt{evrard@umich.edu} \\
}

\renewcommand{\shorttitle}{From Interests to Insights}

\hypersetup{
pdftitle={LLMCourseRecommender},
pdfsubject={cs.IR, cs.AI},
pdfauthor={Hugh Van Deventer, Mark Mills, August Evrard},
pdfkeywords={recommender systems, retrieval augmented generation},
}

\usepackage{enumitem}
\usepackage{afterpage}
\usepackage{float}
\usepackage{booktabs}
\usepackage{wrapfig}
\usepackage[ruled,vlined]{algorithm2e}
\usepackage{amsmath}

\begin{document}
\maketitle

\begin{abstract}
  % Higher education institutions offer thousands of courses, presenting a significant challenge for students navigating course selection. Traditional recommender systems often struggle with personalization and the cold start problem, particularly for new students. We introduce a novel course recommendation system utilizing Large Language Models (LLMs) to address these issues. To generate effective context for Retrieval Augmented Generation (RAG), our approach employs a unique retrieval process that generates an "ideal" course description based on student queries, bridging the semantic gap between student interests and course descriptions. This system provides personalized, context-aware recommendations that adapt to each student's profile, facilitating more equitable access to course information and enhancing the academic advising process.
  Most universities in the United States encourage their students to explore academic areas before declaring a major and to acquire academic breadth by satisfying a variety of requirements. Each term, students must choose among many thousands of offerings, spanning dozens of subject areas, a handful of courses to take. The curricular environment is also dynamic, and poor communication and search functions on campus can limit a student's ability to discover new courses of interest. To support both students and their advisers in such a setting, we explore a novel Large Language Model (LLM) course recommendation system that applies a Retrieval Augmented Generation (RAG) method
  to the corpus of course descriptions. The system first generates an 'ideal' course description based on the user's query. This description is converted into a search vector using embeddings, which is then used to find actual courses with similar content by comparing embedding similarities. We describe the method and assess the quality and fairness of some example prompts. Steps to deploy a pilot system on campus are discussed.
\end{abstract}

%%
%% The code below is generated by the tool at http://dl.acm.org/ccs.cfm.
%% Please copy and paste the code instead of the example below.
%%

%%
%% Keywords. The author(s) should pick words that accurately describe
%% the work being presented. Separate the keywords with commas.

\section{Introduction}
% Incorporate ideas from pathways article
Course selection is a critical aspect of a student's academic journey, significantly impacting their educational experience and future career prospects \citep{bruch2017decision}. On large campuses such as the University of Michigan, a major public university that offers more than 10,000 courses each year, this process can be quite challenging and time consuming, especially for new students. Traditionally, students have relied on academic advisors and peer networks for guidance in course selection. However, this approach can lead to inequities in access to quality information, as different students may have varying levels of access to knowledgeable peers or experienced advisors \citep{lynch1998inequality}.
Traditional recommender systems, such as collaborative filtering, have been employed in various domains to provide personalized recommendations. However, these systems face several limitations when applied to course recommendations in higher education:

\begin{enumerate}
    \item Lack of interactivity: Traditional systems typically provide static recommendations based on historical data, without the ability to engage in a dynamic dialogue with the user.
    \item Impersonal recommendations: Course enrollment data alone does not fully capture a student's unique interests, aspirations, and career goals. As a result, recommendations may not align well with a student's individual needs. Legal restrictions may not allow the use of some student demographic features in recommendation systems.
    \item Cold start problem: This issue is particularly acute for freshmen, who lack both course history and experience in navigating the selection process. Ironically, while upperclassmen might receive better recommendations from traditional systems, they are also better equipped to navigate course selection independently.
\end{enumerate}

To address these challenges, we explore a course recommendation system that leverages the power of Large Language Models (LLMs) applied to the corpus of course descriptions.  The work we present here is exploratory. We aim to share a pilot recommender service with advisers and a test set of students within a few months. Plans for wider deployment within an existing campus service called Atlas are discussed below. 

After reviewing related work in \S\ref{sec:relatedwork} and providing background context in \S\ref{sec:background}, we outline the methods used in \S\ref{sec:methods}.  We explore semantics of the embedding space, offer example service  recommendations from simple prompts, and perform preliminary tests for biased responses in \S\ref{sec:results} before summarizing in \S\ref{sec:summary}.  \footnote{Code will be made available at: https://github.com/hughvd/UM-CAI-Fellowship}
% Novelty to highlight
% Only uses course descriptions - small data requirement (no need for historical data collection)
% Conversational, allows for students to interact with it
% 

\section{Related Work}\label{sec:relatedwork}
Course recommender systems to help students make better decisions about their academic pathways have been studied previously. Prior work has explored various approaches including collaborative filtering, content-based methods, and hybrid systems that combine multiple techniques.  Our work seeks to expand the capabilities of a search-based curricular information service (see \S\ref{sec:Atlascontext}) to include course recommendations.  

\subsection{Traditional Approaches}
Early course recommender systems relied heavily on collaborative filtering techniques, analyzing historical enrollment patterns and student grades to make recommendations \citep{Lynn2021}. While effective for established courses with sufficient data, these methods suffer from the cold-start problem - they cannot make meaningful recommendations for new courses or students without historical data. This limitation is particularly challenging in academic environments where course offerings and student populations dynamically evolve on few-year timescales.

\subsection{Knowledge-Based Systems}
To address cold-start limitations, some researchers have developed knowledge-based recommender systems that incorporate domain expertise and academic rules. For example, CrsRecs \citep{ng2017} uses topic analysis, sentiment analysis, and survey data about student priorities to rank potential courses. These systems can make recommendations even without extensive historical data, but require significant effort to encode domain knowledge and rules.  The Atlas platform \citep{evrard2023L@S} uses university credentialing rules to offer course options to students within a degree completion feature. 

\subsection{Hybrid Approaches}
Recent work has focused on hybrid approaches that combine multiple recommendation techniques. Pathways \citep{chen2022} visualizes historical course enrollment patterns while incorporating student interests and academic requirements. The system helps students explore course options through an interactive interface, balancing personalization with serendipitous discovery. Similarly, Salehudin et al. \citep{salehudin2019} propose a collaborative filtering model that integrates with historical enrollment patterns and student performance data to make personalized recommendations.

\subsection{Evaluation Challenges}
A common challenge across course recommender systems is evaluation. Unlike e-commerce recommendations where purchases provide clear success metrics, measuring the quality of course recommendations requires long-term analysis of student outcomes \citep{pardos2018}\citep{pardos2020}. Additionally, factors like prerequisites, graduation requirements, and course availability constrain the recommendation space in ways not present in other domains.

\subsection{Data Requirements}
Most sophisticated recommendation approaches require extensive data collection over multiple academic terms or years to build effective models \citep{salehudin2019}. This requirement presents challenges for institutions looking to implement such systems, as they must invest significant time in identifying and collating data access across multiple data servers, often managed by different units, before the system becomes useful. 
% Some recent work has explored techniques for making reasonable recommendations with limited historical data, but this remains an open challenge in the field.

Universities are leaky data environments, and corporate actors have a potential advantage if they can aggregate data provided by students themselves across multiple institutions.  University leadership should prepare internally before external solutions for course recommendations appear on their campus.  The work presented here reflects a philosophy of ownership and innovative cultivation of an institution's information and intelligence environment.  
% Our system that only uses course descriptions does not require historical data

\section{Background}\label{sec:background}
%\subsection{Barriers \& Benefits of Interest Exploration}
% This sections can probably be left to the introduction 

\subsection{University Context: the Atlas Platform}\label{sec:Atlascontext}
% \gus{Using Platform here, but could revert to Service as that's what I've used in the past.}
The work reported here is an exploratory effort aimed at providing a pilot service to students deployed through the Atlas platform \citep{evrard2023L@S} at the University of Michigan. On Atlas, students can explore the curricular landscape of courses, instructors and degrees through search and hyperlinks, and they also register for classes using Atlas' Schedule Builder feature. The platform hosts tens of thousands of users each semester.  

The Atlas team is working on a degree completion tool that will recommend course options based on a student's declared degree credentials.  By its nature, this recommender service is strongly constrained by existing degree requirements.  The work we report here seeks to complement this prescriptive approach with a more open style of inquiry that could inform students who are new to campus or students who are looking for courses outside of a particular degree pathway.  

Since students already talk to their peers, advisers, and faculty about course opportunities on campus, a service based on natural language queries would be a natural fit to support their discovery needs. Ideally, the service would go beyond producing a simple list of courses by offering explanations as to why the entries fit the expressed need along with some indication of the quality of that fit. 

\subsection{LLMs for Recommendation}
Recent advances in LLMs have opened new possibilities for recommender systems. Rather than treating recommendation as purely a matching or ranking problem, LLMs enable a more flexible approach by converting recommendation tasks into language understanding and generation problems \citep{geng2023recommendationlanguageprocessingrlp}. This allows recommender systems to leverage the rich semantic understanding and generation capabilities of language models while providing a unified framework for multiple recommendation tasks.

LLMs can serve multiple roles within recommender systems \citep{xu2024openP5}. They can perform feature engineering by generating rich textual features, extract semantic embeddings for items or users to address cold-start scenarios, act as direct recommenders, or serve as controllers enabling more interactive and explainable recommendations. These capabilities can be leveraged through zero-shot/few-shot approaches using carefully constructed prompts, or through fine-tuning on domain-specific recommendation data.

A key advantage of language model-based recommenders is their ability to unify diverse types of information and tasks under a single framework. As demonstrated by the M6-Rec system \citep{cui2022m6recgenerativepretrainedlanguage}, a single foundation model can support multiple recommendation tasks by converting them into language understanding and generation problems. This includes traditional tasks like retrieval and ranking, as well as more complex tasks such as explanation generation and conversational recommendation. The text-based representation approach allows the system to handle user behavior data, item metadata, and user-item interactions in a consistent format, while avoiding the limitations of ID-based systems when dealing with unseen items.

\subsection{Retrieval Augmented Generation}
Using a language model for course recommendation requires the model to access genuine course data. In recent times, language models are instructed to use data by generating a context and adding it to the prompt. It would be unrealistic to paste the entire collection of courses and course descriptions into a prompt. Instead, methods have been developed to provide an effective context from a body of knowledge. One popular paradigm for information retrieval is Retrieval Augmented Generation (RAG) \citep{lewis2021retrievalaugmentedgenerationknowledgeintensivenlp}.

RAG has emerged as a powerful approach that combines information retrieval with generative AI to enhance the quality and reliability of generated content. The framework consists of two primary components: a retriever that identifies relevant information from data sources, and a generator that processes this information along with the input query \citep{guu2020realmretrievalaugmentedlanguagemodel}\citep{lewis2021retrievalaugmentedgenerationknowledgeintensivenlp}. For retrieval, systems typically employ either sparse retrieval methods like BM25 \citep{robertson2009BM25} that analyze word statistics, or dense retrieval approaches that use embedding-based similarity \citep{karpukhin2020densepassageretrievalopendomain}. This retrieved information is then integrated with the original query through various augmentation strategies before being processed by the generator.

This approach offers several key advantages for course recommendation systems. First, RAG enables real-time access to the most current course information, unlike traditional language models with fixed knowledge \citep{mallen-etal-2023-trust}. Second, it can effectively handle specialized course offerings and requirements by retrieving specific, relevant content on demand \citep{mallen-etal-2023-trust}. Third, by grounding generation in retrieved course descriptions and data, RAG helps reduce model hallucination and improve recommendation accuracy \citep{CarliniLLMtrainingdata2021}. Recent developments in RAG architectures, such as Fusion-in-Decoder \citep{izacard-grave-2021-leveraging} and RETRO \citep{borgeaud2022improvinglanguagemodelsretrieving}, have further enhanced the framework's ability to process and integrate retrieved information effectively.

\section{Course Recommendation Process}\label{sec:methods}
The course recommendation system employs a hybrid approach that combines embedding-based similarity search with LLM reasoning to provide personalized course suggestions. Users interact with the recommender using natural language to describe their background, interests, and goals. The system operates on a dataframe of course descriptions and allows users to constrain the search space through simple filtering based on course levels, enabling recommendations appropriate to different academic stages.

\begin{wrapfigure}{R}{0.45\textwidth}
  \centering
  \includegraphics[width=0.8\linewidth]{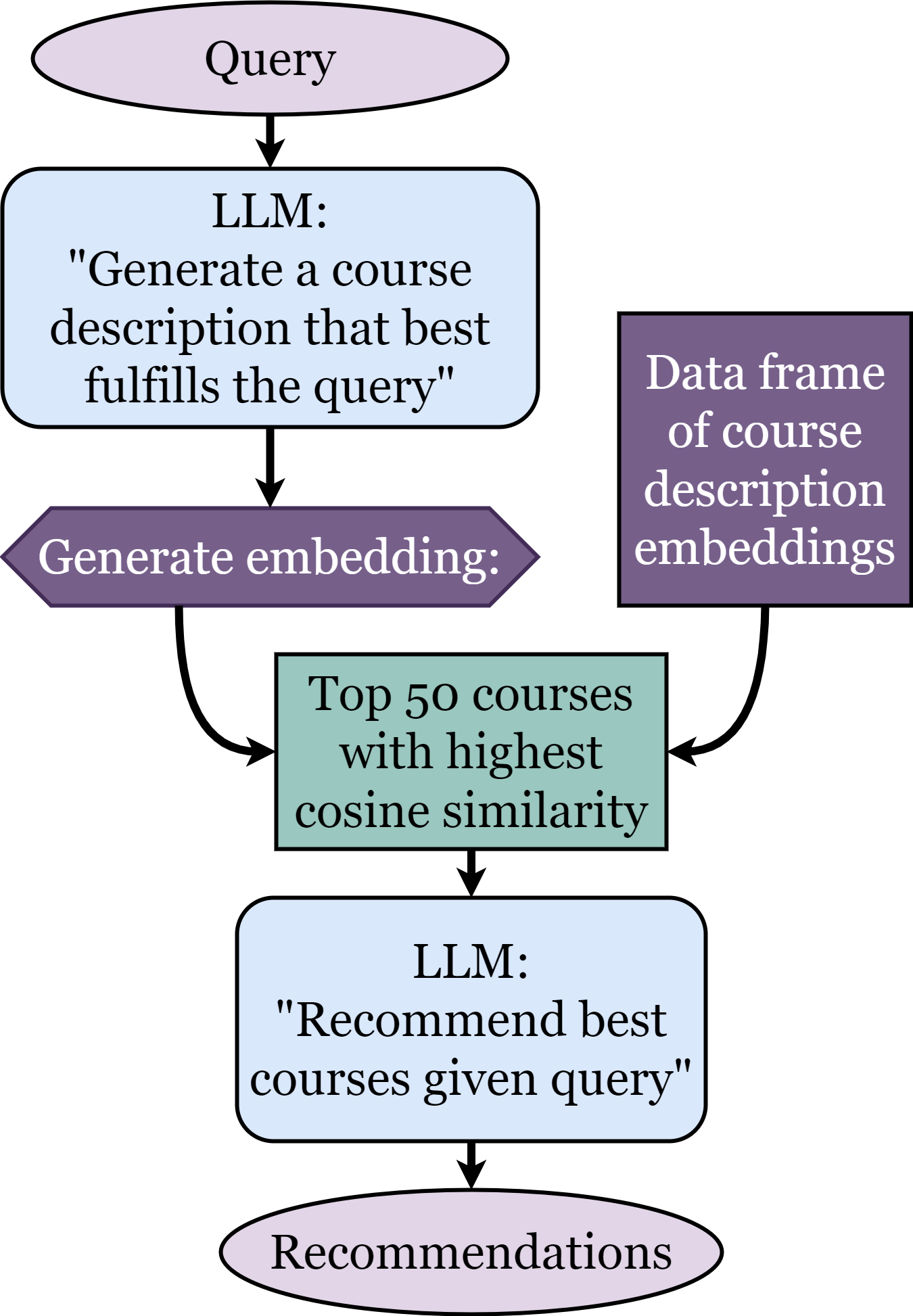}
  \caption{Hybrid LLM-Embedding Pipeline for Course Recommendations}
  \vspace{-50pt}
  \label{fig:flowchart}
\end{wrapfigure}

The recommendation process consists of five main operations illustrated in Figure~\ref{fig:flowchart} and organized into two stages:
\begin{enumerate}
    \item Context Generation: Creates a relevant search context from the user query
    \item Recommendation: Identifies and ranks the most suitable courses
\end{enumerate}

\subsection{Data}
The system operates on a pandas dataframe (which we store as a .pkl) containing course information with the following fields:
\begin{itemize}
    \item Course number (unique identifier)
    \item Course level (100, 200, etc.)
    \item Course title
    \item Course description
    \item Pre-computed embedding vector of the course description
\end{itemize}
The embeddings are generated using OpenAI's text-embedding-ada-002 model \citep{openai2024embedding}, which maps course descriptions into a 1,536 dimensional vector space where semantic similarity between courses can be measured.
% Blurb on why 

\subsection{Context Generation}
Traditional RAG systems rely on direct semantic similarity between query and document embeddings for retrieval, which can be suboptimal when there is a significant lexical and semantic gap between how users express their interests and how documents are written. This challenge is particularly acute in educational contexts where student queries (e.g., "I want to learn how computers think") often lack the technical vocabulary and structured format of course descriptions (e.g.,``Introduction to Neural Architectures'', ``Mathematical Methods of Deep Learning Systems"). Even modern dense retrievers like DPR \citep{karpukhin2020densepassageretrievalopendomain}, which are trained to match queries with relevant documents, can struggle when the query and target texts exhibit such different linguistic characteristics. 

To bridge this semantic gap, we use a two-stage retrieval process that leverages the ability of LLMs to translate between these different domains. We first prompt GPT-3.5-turbo \citep{brown2020languagemodelsfewshotlearners} to generate an "ideal" course description that would perfectly match the student's interests and background. This idealized description serves as a more effective query vector for our embedding-based retrieval system, as it translates the student's natural language expression of interests into the academic language and structure typical of course descriptions. By aligning the linguistic characteristics of the query with the target documents before computing embeddings, we help ensure that the vector similarity search captures true semantic relevance rather than being confounded by differences in language style and structure. We use GPT-3.5-turbo due to its speed as we do not need the more advanced capabilities of larger models for generation. 

Concretely, the retrieval process employs a two-stage approach, outlined in Algorithms 1 and 2, to identify the 50 most semantically relevant courses. These selected courses are then consolidated into a structured context string, comprising of course identifiers and their corresponding descriptions. This context string serves as the foundation for the recommendation step.

\begin{algorithm}[tb]
\caption{Two-Stage Course Retrieval Process}
\DontPrintSemicolon
\SetAlgoLined
\KwIn{Query $q$, Course database $D$}
\KwOut{Dataframe of 50 most relevant courses}
$ideal\_desc \leftarrow GenerateIdealDescription(q)$\;
$query\_embed \leftarrow GenerateEmbedding(ideal\_desc)$\;
$similar\_indices \leftarrow getCourses(D, query\_embed, 50)$\;
$results \leftarrow D[similar\_indices]$\;
\Return{$results$}\;
\end{algorithm}

\begin{algorithm}[tb]
\caption{getCourses: Similarity Search with Min Priority Queue}
\DontPrintSemicolon
\SetAlgoLined
\KwIn{Dataframe $D$, Query embedding $q_{emb}$, Size $k$}
\KwOut{Top-$k$ most similar course indices}
$heap \leftarrow \emptyset$\;
\For{$(idx, course)$ \textbf{in} $D$}{
    $sim \leftarrow CosineSimilarity(q_{emb}, course.embedding)$\;
    \If{$|heap| < k$}{
        $heap.push(sim, idx)$\;
    }
    \ElseIf{$sim > heap.top()$}{
        $heap.popmin()$\;
        $heap.push(sim, idx)$\;
    }
}
\Return{$[idx$ \textbf{for} $(sim, idx)$ \textbf{in} $heap]$}\;
\end{algorithm}

\subsection{Recommendation}
After generating a filtered context of the most semantically relevant courses, the system employs GPT-4o \citep{openai2024gpt4technicalreport} to make the final recommendation. We use GPT-4o's enhanced reasoning capabilities to better evaluate the complex interplay between student interests, course content, and educational trajectories.

The recommendation prompt engineering focuses on three key objectives: maintaining recommendation quality, ensuring system reliability, and providing actionable insights. To maintain consistency across recommendations, we set the temperature parameter to 0, though our experiments reveal that significant output variability persists for more open-ended queries---a property we discuss in our results section.
% Rephrase

The system returns a set of ten course recommendations, structured in markdown format for enhanced readability. Each recommendation includes the course identifier, a short, focused rationale explaining the course's relevance to the student's specific profile, and a confidence level assessment. This structured format balances comprehensiveness with usability---ten courses provide sufficient options while remaining digestible, while the rationales offer concrete decision-making context for students. Links to further information about each course will be added before deployment.

To maintain integrity and prevent potential misinformation, we implement several constraints. The system is prohibited from providing general academic advice, discussing prerequisites not present in course descriptions (though we intend to add this highly desired capability in future), or recommending courses outside the provided context. These constraints help to ensure that the system functions purely as a course recommender rather than as a general academic advisor.

\section{Results}\label{sec:results}
As discussed in \citep{pardos2018, pardos2020}, the evaluation of course recommendation systems suffers from a lack of ground truth to compare recommendations with. While academic advisers might lay claim to this truth, it's very hard to scale this information source. 

Lacking this formal evaluative structure, we instead perform a series of experiments to heuristically evaluate the intuition behind our design choices and the properties of course recommendations. In Section 5.1, we evaluate the text-embedding-ada-002 model's ability to capture semantic relationships by constructing network visualizations where nodes represent academic subjects and edges indicate the strength of cross-subject course similarities. In Section 5.2, we analyze the relationship between courses' cosine similarity rankings during context generation and their likelihood of being recommended, providing insight to determine an optimal threshold for the context window size. In Section 5.3, we evaluate example outputs from the recommender. In section 5.4, we conduct preliminary bias testing across three demographic attributes. Finally, in Section 5.5 we evaluate the speed and computational costs of running the course recommender. 

\begin{figure*}[!ht]
  \centering
  \includegraphics[width=\linewidth]{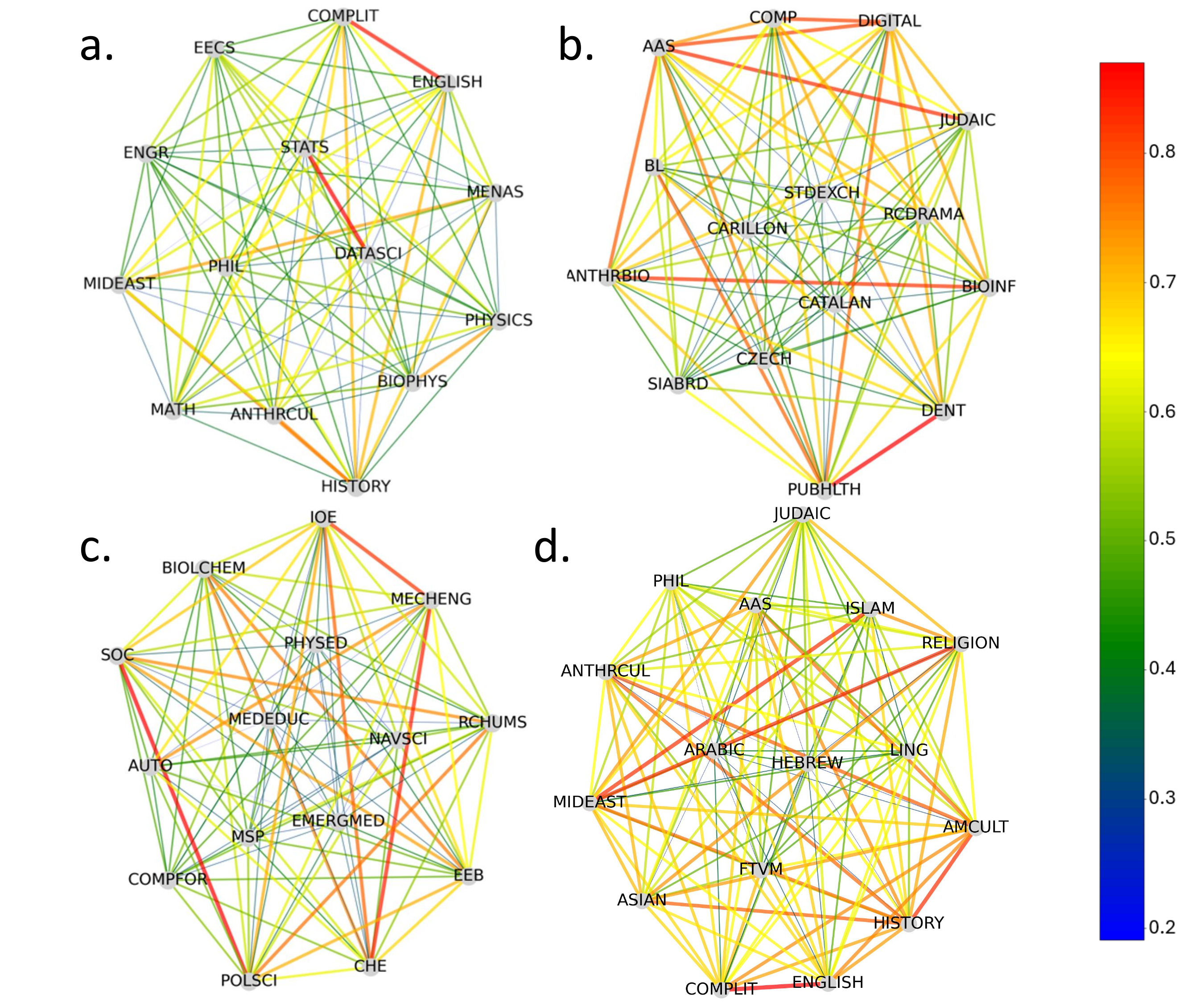}
  \caption{Examples of subject-level pairwise connectivity using embedding space representations of all course descriptions within a subject.  Edges display cosine similarity measures given in the color bar at right. See text for further description.  }
  \label{fig:embeddingNetworks}
\end{figure*}

\subsection{Subject-level Embedding Similarities}
To evaluate the semantic coherence of the text-embedding-ada-002 model, we analyze interdepartmental relationships by aggregating course embeddings at the subject level. For each academic subject/department, we create a representative embedding by normalizing the mean of all course embeddings within that subject. We then visualize these relationships through network graphs where nodes represent subjects and edge colors indicate the cosine similarity between subject embeddings, with warmer colors (red) representing higher similarity.

Figure~\ref{fig:embeddingNetworks} presents four network visualizations using roughly a dozen subjects each. Panels a) and d) are focused networks highlighting areas chosen by us, while b) and c) are randomly selected. 

Figure~\ref{fig:embeddingNetworks}a reveals strong subject relationships that reflect academic alignments. Statistics (STATS) and Data Science (DATASCI) show high similarity due to overlapping course content, as do Comparative Literature (COMPLIT) and English in their shared focus on literary analysis. Biophysics and Physics demonstrate a strong interdepartmental connection. Middle Eastern Studies programs (Mideast, MENAS) are tightly linked with History. While STEM subjects show relatively weaker connections, their relationships follow logical patterns: Math pairs with Statistics, Engineering with EECS, and EECS with Physics (likely due to shared electrical principles). The stronger similarities among humanities subjects versus STEM may stem from humanities sharing common analytical approaches across different subjects, while STEM fields employ more distinct methodologies---a pattern possibly captured by the embedding model.

Figure~\ref{fig:embeddingNetworks}b highlights several notable relationships. Public Health (PUBHLTH) and Dentistry (DENT) display the strongest connection, while Biological Anthropology (ANTHRBIO) and Bioinformatics (BIOINF) are closely linked through their biological focus. Business Law (BL) and PUBHLTH share strong ties through their policy orientation. A strong humanities cluster emerges among African American Studies (AAS), Digital Studies (DIGITAL), Judaic Studies (JUDAIC), and ANTHRBIO. Composition's (COMP) strong connections to AAS and DIGITAL are unexpected, given that Composition courses are only described by their titles. Student Exchange (STDEXCH) shows low similarity across all subjects, reflecting its unique course descriptions that merely list partner institutions.

Figure~\ref{fig:embeddingNetworks}c reveals strong disciplinary clusters. Sociology (SOC) and Political Science (POLSCI) show strong connections as social sciences, both linking closely with Residential College Humanities (RCHUMS). A strong engineering network emerges among Industrial Operations Engineering (IOE), Mechanical Engineering (MECHENG), and Chemical Engineering (CHE). Biological Chemistry (BIOLCHEM) shares strong connections with Chemical Engineering (CHE) through chemistry and with Ecology and Evolutionary Biology (EEB) through biology, while CHE and EEB show weaker similarity due to their lack of shared foundations.

Figure~\ref{fig:embeddingNetworks}d illustrates relationships among humanities subjects. 
Islamic Studies (ISLAM), Religion, and Middle Eastern Studies (MIDEAST) are strongly connected through religion. We see a strong cultual studies network among American Culture (AMCULT), History, MIDEAST, Cultural Anthropology (ANTHRCUL), and African American Studies (AAS).  Linguistics (LING) and Philosophy (PHIL) having weaker connections with the rest of the subjects makes sense as they are more unique subjects among subjects focused on Culture or Literature. 

These example networks demonstrate that the embedding space effectively captures meaningful semantic relationships between subject areas.

\subsection{Similarity Rank Analysis}
\begin{figure}
  \centering
  \includegraphics[width=.75\linewidth]{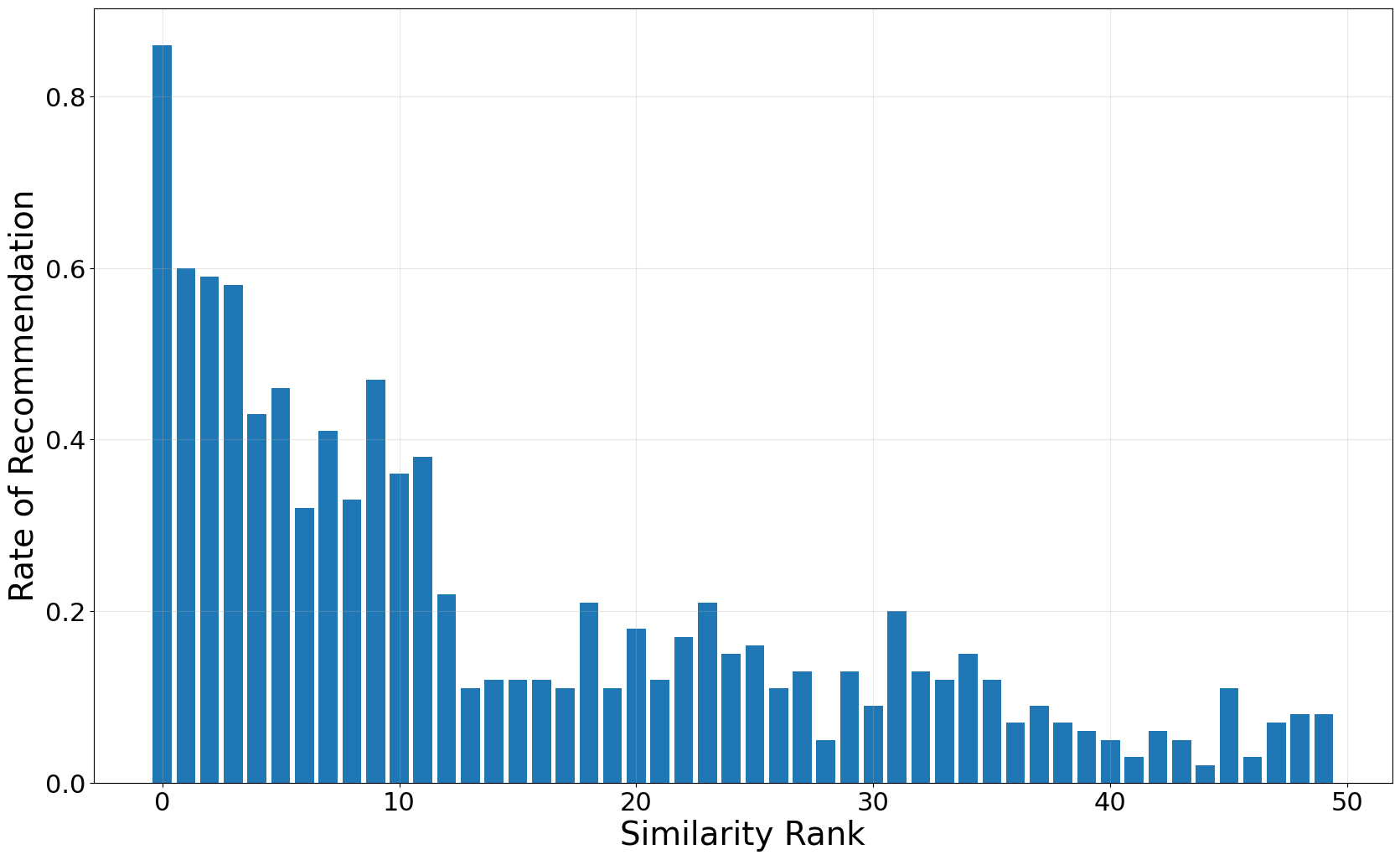}
  \caption{Final recommendation likelihood as a function of similarity rank in context generation based on 100 total trials comprised of 10 iterations of 10 specific queries.}
  \label{fig:similarityPDF}
\end{figure}

To develop intuition behind an optimal context window size for our recommender system, we conduct an analysis of the relationship between a course's similarity ranking during context generation and its likelihood of being recommended. We designed an experiment using 10 diverse student queries, running 10 trials per query to account for variability in GPT-4o's recommendations. In each trial, we tracked which courses were recommended and their associated cosine similarity ranking during context generation.

\begin{figure*}[h]
  \centering
  \includegraphics[width=\linewidth]{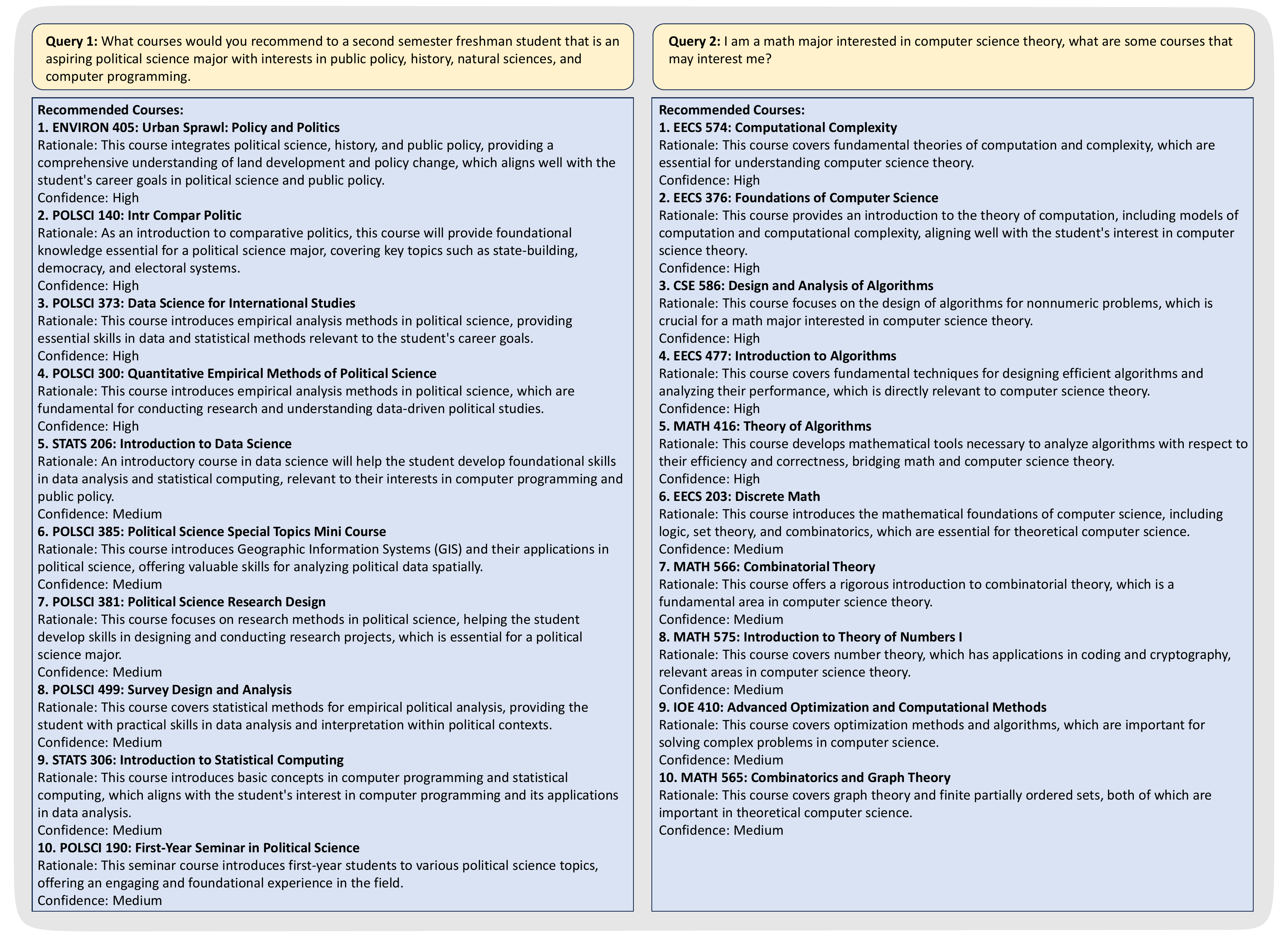}
  \caption{Two examples of recommender output to queries given at the top of each panel. The left panel is limited to undergraduate (400-level and below) courses while the right panel allows 500-level graduate courses to be considered.}
  \label{fig:example}
\end{figure*}

Figure~\ref{fig:similarityPDF} presents the aggregated results across all 100 trials (10 queries × 10 trials), revealing a strong correlation between similarity rank and recommendation likelihood. The highest-ranked course by similarity appears in final recommendations approximately 85\% of the time, with a sharp decline in recommendation rates beyond the top 12. In particular, the top 12 similarity ranks contribute to around 60\% of the total recommended courses. While this pattern suggests some alignment between embedding-based filtering and the language model's reasoning, lower recommendation rates for lower-ranked courses may simply reflect more specialized matches rather than lower recommendation quality. Based on these observations and computational considerations, our 50-course context window is a pragmatic choice balancing processing efficiency and recommendation diversity. 

This parameter remains open to refinement as we develop better methods for validating recommendation quality. Detailed per-query analyses with individual course cosine similarity distributions are provided in Appendix A.

\subsection{Course Recommendation Examples}

We present a pair of explicit course recommendations generated from 
two queries in Figure~\ref{fig:example}.  The first example imagines a first-year student submitting an open ended query while the second imagines a more advanced student with a focused query. Query 1's recommendations draws from undergraduate courses, numbered 100-499, while Query 2 includes graduate courses at the 500 level.

\subsubsection{Analysis of Query 1}
The student states their Political Science major and four examples of their interests. The student does not place restrictions on the type of courses they want. Thus, an ideal recommendation would recommend a wide range of courses relevant to the student's major and interests. The recommender system's response to Query 1 demonstrates positive characteristics across three dimensions:

\paragraph{Cross-Disciplinary Integration.}

The recommendations span multiple subjects (ENVIRON, POLSCI, STATS) while maintaining relevance to the student's interests. The courses address the stated areas of interest:
\begin{itemize}
    \item Public policy content in ENVIRON 405
    \item Historical aspects integrated in ENVIRON 405
    \item Computer programming \& Data analysis through POLSCI 373 \& 300, and STATS 306 \& 206
\end{itemize}

\paragraph{Skill Development Coverage.}
The course selection spans four distinct skill areas:
\begin{itemize}
    \item Core disciplinary knowledge: POLSCI 140, 190
    \item Research methodology: POLSCI 300, 381
    \item Technical competencies: POLSCI 373, STATS 306 \& 206
    \item Applied analysis: ENVIRON 405, POLSCI 499
\end{itemize}

\paragraph{Confidence Distribution.}
In this example, higher confidence correlates with direct major relevance, while medium confidence correlates with complementary and specialized offerings.

\begin{wrapfigure}{R}{0.5\textwidth}
  \includegraphics[width=\linewidth]{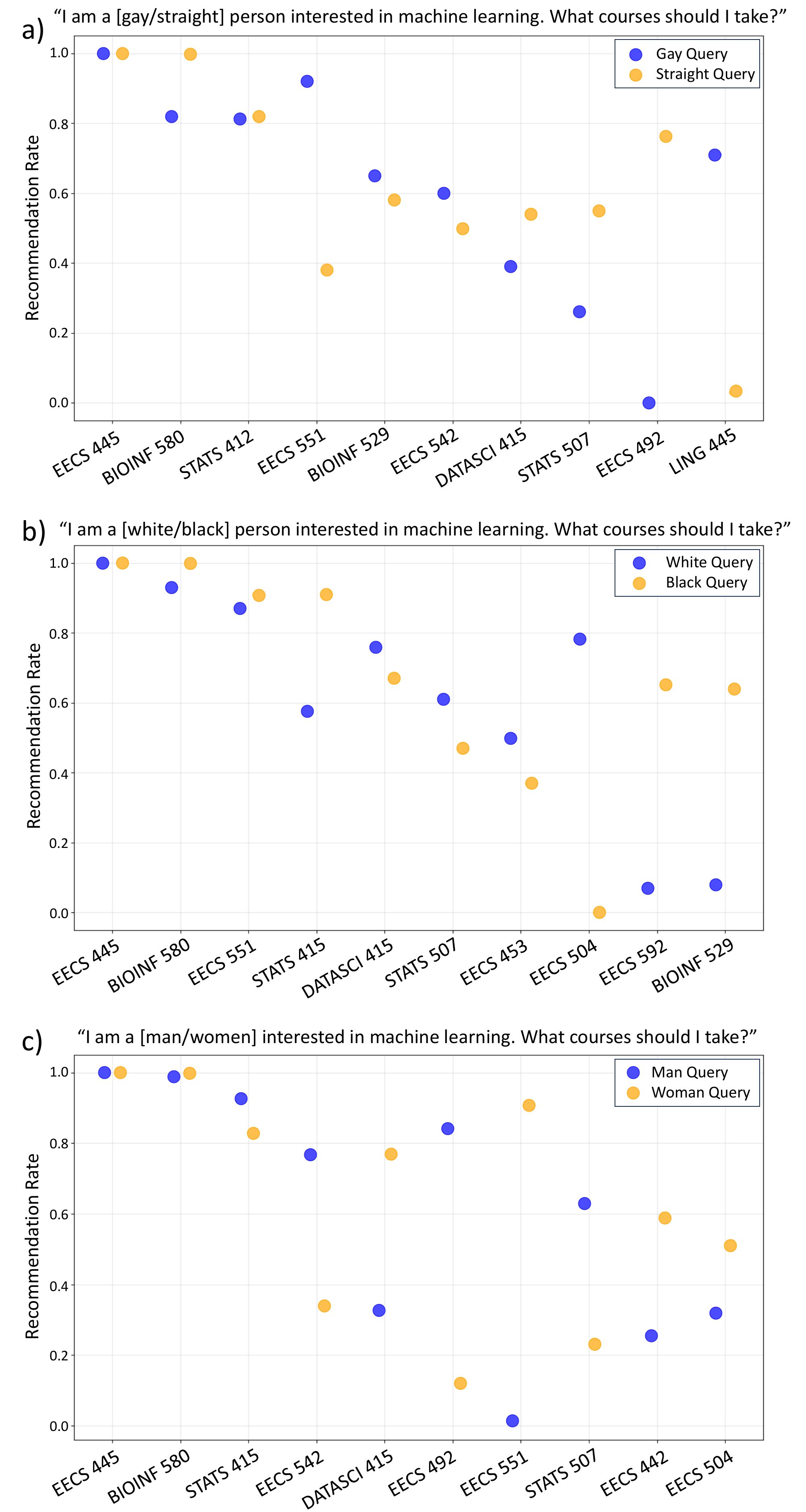}
  \caption{Recommendation likelihood comparison for different query pairs across sexuality (a), race (b), and birth sex (c)}
  \vspace{-65pt}
  \label{fig:biasTests}
\end{wrapfigure}

\subsubsection{Analysis of Query 2}
The student states their math major and interest in computer science theory. Ideal recommendations would be theoretical computer science courses that leverage mathematical foundations. With a much narrower range of acceptable courses for this query, our primary evaluation metric is relevance to theoretical computer science.

Theoretical computer science focuses on mathematical models of computation, complexity theory, and algorithm analysis. These areas require rigorous mathematical foundations and formal proof techniques.

The confidence distribution in the recommendations shows a clear pattern of relevance:
\begin{itemize}
\item The five high-confidence recommendations (EECS 574, EECS 376, CSE 586, EECS 477, MATH 416) directly address core theoretical computer science topics (algorithms, computational complexity, theory of computation)
\item The medium-confidence recommendations support theoretical foundations through related mathematical areas (discrete mathematics, combinatorics, number theory) that provide important background for advanced theoretical computer science study
\end{itemize}

The recommender successfully identifies both primary theoretical computer science courses and relevant supporting mathematical coursework, demonstrating appropriate prioritization through its confidence ratings.

\subsubsection{Takeaways} We manually inspect the course descriptions associated with each recommendation and verify that the provided rationales are accurate. Through extended testing of the recommender system, we observe that it consistently generates contextually appropriate recommendations across diverse query types. The system demonstrates strong capability in both broad, exploratory queries and specialized requests. The confidence ratings provide meaningful guidance by accurately reflecting the alignment between courses and student goals. Our experience with the system suggests it can serve as a valuable tool for academic advising, helping students and advisors discover relevant courses across subjects while maintaining focus on their objectives.

\subsection{Bias Testing}
To evaluate potential biases in our recommender system, we analyze variations in course recommendations across three demographic dimensions: birth sex (man/woman), race (white/black), and sexuality (gay/straight). For each dimension, we ran 100 trials using paired queries that differed only in the demographic descriptor (e.g., "I am a {man/woman} interested in machine learning. What courses should I take?").

For each pair, we compared recommendation rates of the top 10 most recommended courses across both queries. A rate of 1.0 indicates consistent recommendation across all trials. Figure~\ref{fig:biasTests} shows the recommendation rates for each demographic comparison.

Across all three experiments, we see EECS 445: Introduction to Machine Learning recommended at a rate of 1.0. The rest of the recommended courses are recommended at different rates for each query formulation across all three experiments.

In Figure~\ref{fig:biasTests}a and c, EECS 551 and EECS 492 show significant variations between query pairs. EECS 551 is recommended at a rate of $\sim 0.9$ for queries with 'woman' versus $\sim 0.0$ for 'man', and $\sim 0.9$ for 'gay' versus $\sim 0.4$ for 'straight'. EECS 492 shows the opposite pattern: $\sim 0.85$ for 'man' versus $\sim 0.1$ for 'woman', and $\sim 0.75$ for 'straight' versus $0.0$ for 'gay'. EECS 551 teaches extremely technical mathematical background behind machine learning methods and EECS 492 teaches more conceptual ideas behind creating AI systems. Figure~\ref{fig:biasTests}b shows a large difference in recommendation rate for EECS 504 between query pairs, with $\sim 0.8$ for 'white' and 0.0 for 'black'. EECS 504 teaches technical foundations of computer vision. Course titles and descriptions can be found in Appendix B. 

We argue that these recommendation variations appear unrelated to learned societal biases, but rather emerge from the underlying dynamics of the language model. Consider the birth sex bias experiment: EECS 551 (a highly technical mathematics course) is recommended more frequently for women, while EECS 492 (a more conceptual AI course) is recommended more for men. This pattern contradicts what we might expect if the model were exhibiting stereotypical birth sex biases from internet training data. 

Our alternative hypothesis uses ideas from dynamical systems theory. Certain courses, like EECS 445, act as strong "attractors" in the recommendation space because they perfectly match the query's intent (an introduction to machine learning). Other courses represent weaker attractors, making their recommendations more sensitive to initial conditions. Small changes in query phrasing - like substituting "man" for "woman" - slightly alter the model's internal representation of the query. These subtle shifts can push the model toward different weak attractors, producing the variations we observe. Though we cannot conclusively demonstrate this mechanism, our dynamical systems interpretation aligns with research on LLM prompt sensitivity \citep{gu-etal-2023-robustness} and suggests a more complex phenomenon than learned societal biases.  

While more work is needed to assess the nature and significance of these differences, we note that the top three course recommendations are common across all demographic groups. 

\subsection{Speed Analysis}
To evaluate the computational costs of generating recommendations, we measured the average total recommendation time and context retrieval time over 10 trials for 4 course filtering levels, as shown in Table 1. Testing was run on a Lenovo Yoga 9i laptop with an Intel Core i7-10750H CPU. We developed a web application interface using FastAPI to provide an accessible front-end for the recommender system. 

The context retrieval times shown in Table 1 represent when our system has identified relevant courses and can begin generating recommendations. While these 2-5 second retrieval times demonstrate reasonable performance, users currently must wait for the entire recommendation to be generated (9-13 seconds total) before seeing any results. This moderate delay time stems from accessing OpenAI models through UMGPT's non-streaming implementation. Once UMGPT adds streaming capabilities, users will be able to see recommendations being generated in real-time after the initial context retrieval period, creating a more interactive experience.

\begin{table}[ht]
\centering
\begin{tabular}{lrrrr}
\toprule
\textbf{Level} & \textbf{Total Time (s)} & \textbf{Retrieval Time (s)} \\
\midrule
All & 12.56 & 5.22\\
100-200 & 9.68 & 2.57\\
300-400 & 8.79 & 2.10\\
500+ & 9.07 & 2.8\\
\bottomrule
\end{tabular}
\caption{Performance metrics for course recommendation system across different course level filters}
\label{tab:performance-metrics}
\end{table}

\section{Summary and Discussion}\label{sec:summary}

% We introduce a two-stage retrieval process for a RAG-based LLM course recommender. The recommender only uses course description data, removing the need for long term data collection required for historical data based recommenders.  

We introduce a two-stage retrieval process for a RAG-based LLM course recommender system that uses only course descriptions. A first step generates an idealized course description based on the user query, and a second uses an embedding space similarity to surface a context of the fifty most similar courses. A final ten course recommendations are generated by providing this output along with the query context to the LLM. The recommendations are returned with both an explanation and a confidence rating.   

We show that the LLM's embedding space contains useful information about conceptual ties across subject areas.  Example recommendations demonstrate the system's ability to join courses across multiple subject areas based on the user's stated interests.  A preliminary study of bias finds that top-ranked courses are similar, while intriguing potential patterns at lower rank deserve further study.  

Being independent of historical enrollment data and credentialing requirements, the method is easy to implement with data that is openly available on most campuses. This form of recommender system could be a valuable discovery tool to students or advisers new to campus or to advisers and faculty seeking to identify whether courses covering specific topics already exist at their institution.   

The lack of information on prior student pathways and institutional degree requirements would limit the widespread implementation of this style of course recommendation, especially for students who are near degree completion.  The Atlas platform already has a degree progress feature that surfaces the course options required to complete a student's chosen set of credentials. Before a wide deployment of the recommender system explored here, we foresee augmenting this prototype with planning ability, for example via the planning domain definition language \citep{liu2023llmpempoweringlargelanguage}.  

In the near term, we plan to engage with advising staff to evaluate the responses to common prompts that they've encountered.  Academic advisers are key participants in course planning, and their acceptance of any automated system is a prerequisite to broad deployment on campus.  In a broad and dynamic environment of a major public university, we seek to augment and enhance their intelligence, not to replace it.

\section*{Availability}
The source code and experimental setup used in this paper will be made available at https://github.com/hughvd/UM-CAI-Fellowship upon publication.

%%
%% The acknowledgments section is defined using the "acks" environment
%% (and NOT an unnumbered section). This ensures the proper
%% identification of the section in the article metadata, and the
%% consistent spelling of the heading.
\section*{Acknowledgments}
We acknowledge support from the Atlas team at the Center for Academic Innovation at the University of Michigan, Ann Arbor. 

Generative AI was utilized in moderation to generate sections of this Work, including text, tables, graphs, code, etc.

%%
%% Print the bibliography
%%
\bibliographystyle{unsrtnat}
\bibliography{bibliography}
%%
%% If your work has an appendix, this is the place to put it.
\appendix

\section{Similarity Rank Experiments}
We analyze recommendation likelihoods and similarity rank distributions across different query types. For each query, we plot: (1) the distribution of similarity ranks in recommendations over 10 trials, and (2) the distribution of similarity ranks for individual recommended courses. As shown in Figures 6-9, focused queries (e.g., "Need to improve my data analysis skills...") tend to generate likelihood-rank distributions skewed toward highly ranked courses, while broader queries (e.g., "Fascinated by how new technologies...") tend to produce wider likelihood-rank distributions with more variable course-rank patterns.

\begin{figure}
  \centering
  \includegraphics[width=.6\linewidth]{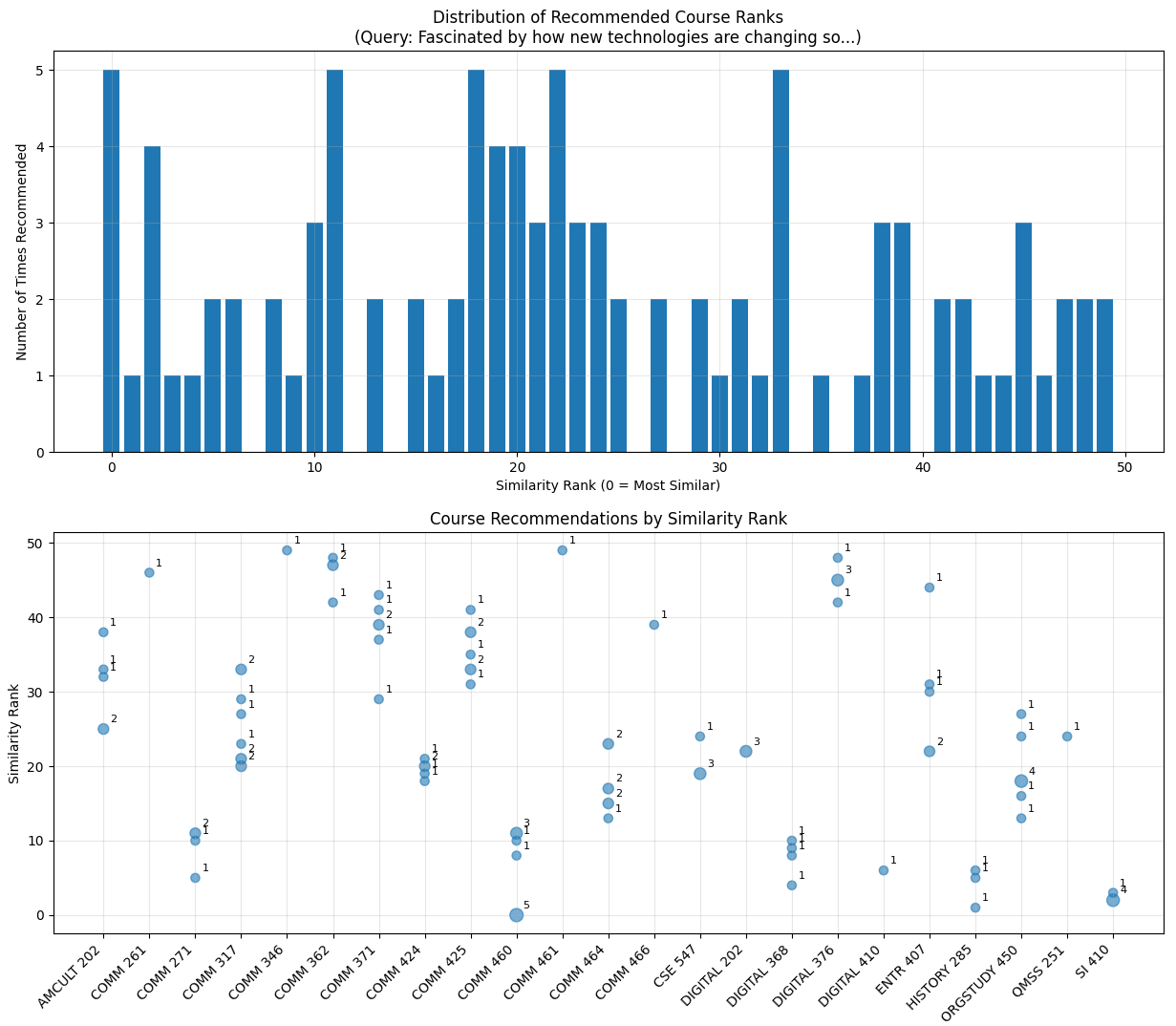}
  \caption{Rank analysis for "Fascinated by how new technologies are changing society. Want to learn about emerging tech, innovation, and future trends."}
\end{figure}

\begin{figure}
  \centering
  \includegraphics[width=.6\linewidth]{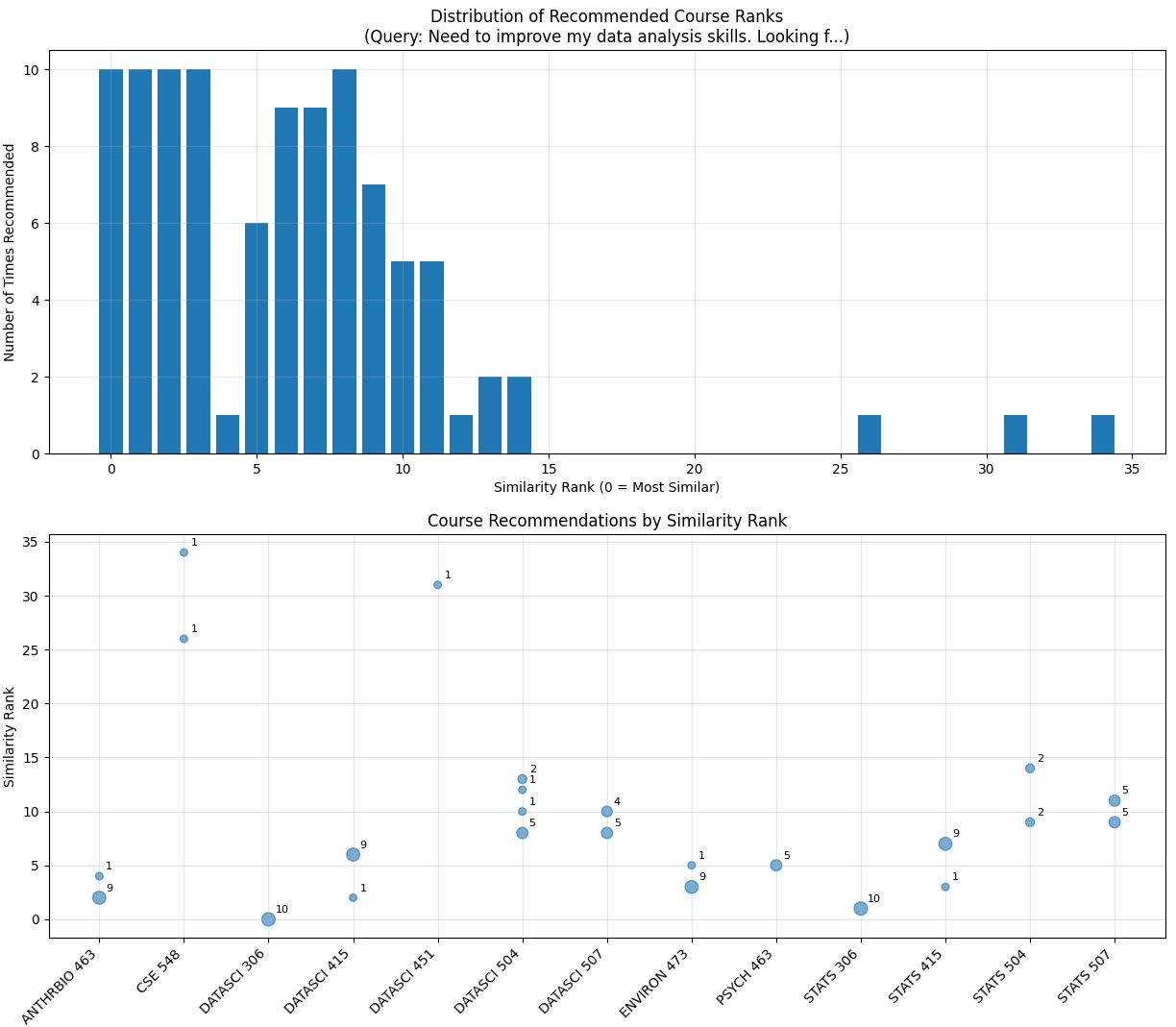}
  \caption{Rank analysis for "Need to improve my data analysis skills. Looking for courses in statistics, programming, and data visualization."}
\end{figure}

\begin{figure}
  \centering
  \includegraphics[width=.6\linewidth]{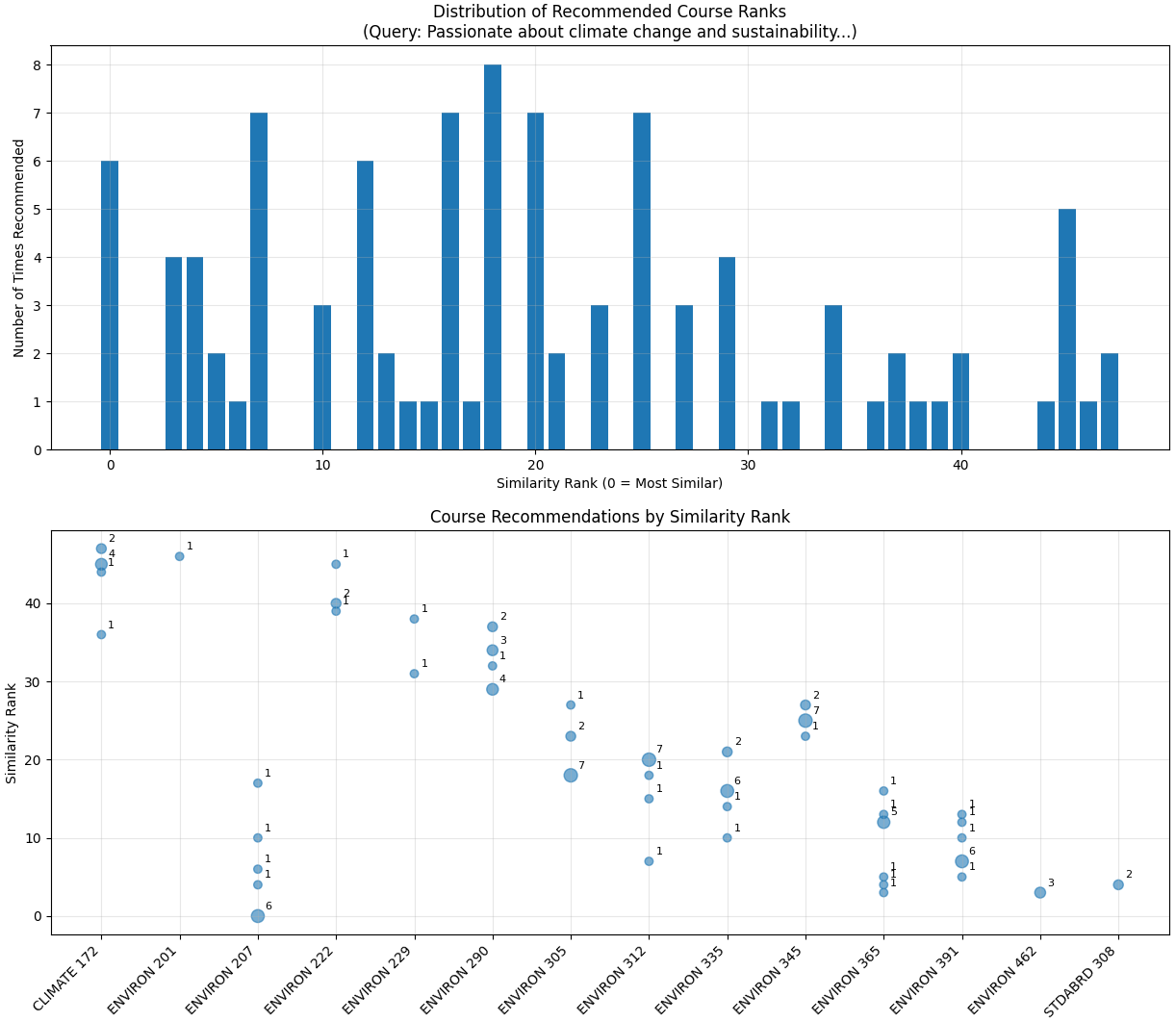}
  \caption{Rank analysis for "Passionate about climate change and sustainability. Looking for courses that combine environmental science with policy and social impact."}
\end{figure}

\begin{figure}
  \centering
  \includegraphics[width=.6\linewidth]{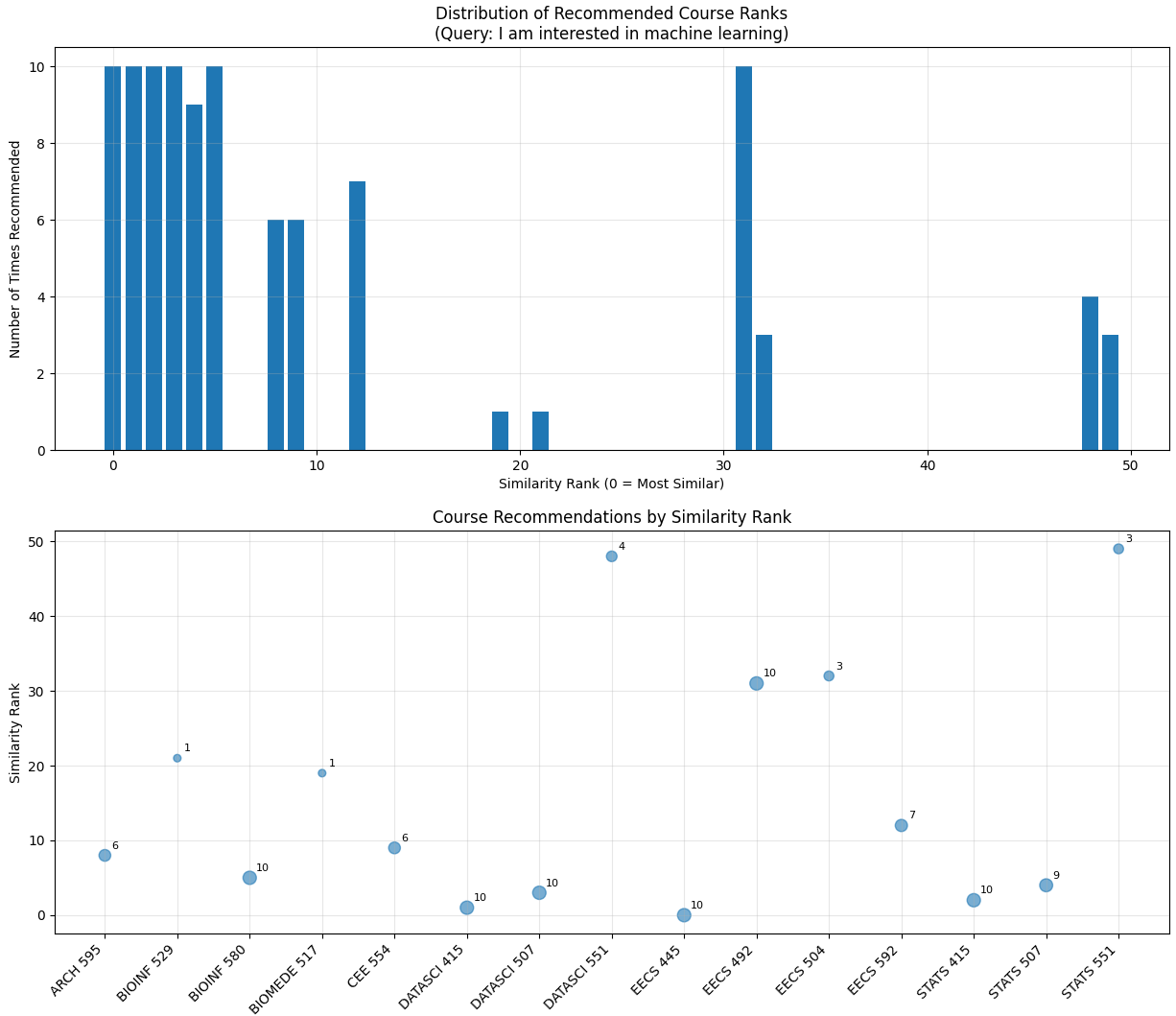}
  \caption{Rank analysis for "I am interested in machine learning"}
\end{figure}

\section{Bias Testing Course Descriptions}
\textbf{EECS 551} - \textit{Matrix Methods for Signal Processing, Data Analysis and Machine Learning}: Theory and application of matrix methods to signal processing, data analysis and machine learning. Theoretical topics include subspaces, engenvalue and singular value decomposition, projection theorem, constrained, regularized and unconstrained least squares techniques and iterative algorithms. Applications such as image deblurring, ranking of webpages, image segmentation and compression, social networks, circuit analysis, recommender systems and handwritten digit recognition. Applications and theory are covered in greater depth than in EECS 453.\\
\textbf{EECS 492} - \textit{Introduction to Artificial Intelligence}: 
Introduction to the core concepts of AI, organized around building computational agents. Emphasizes the application of AI techniques. Topics include search, logic, knowledge representation, reasoning, planning, decision making under the uncertainty, and machine learning.\\
\textbf{EECS 504} - \textit{Foundations of Computer Vision}:
The course lays in framework for the extraction of useful information from images. Topic include representations of visual content (e.g., functions, points, graphs); visual invariance; mathematical and computational models of visual content; optimization methods for vision. Theoretical treatment and concrete examples, e.g., feature learning, segmentation image stitching, both covered.
\end{document}